# Using satellite image classification and digital terrain modelling to assess forest species distribution on mountain slopes – a case study in Varatec Forest District

Ionuț Barnoaiea

„Stefan cel Mare" University of Suceava, Forestry Faculty, Romania
Corresponding author e-mail address: ibarnoaie@usv.ro (I. Barnoaiea)

**Abstract:** The relation between ecological conditions and geomorphological factors is considered the basis for species distribution in Romania. In this context, the location of each species within parts of the mountain slopes is difficult on a medium to brad scale level. The paper presents methodology to combine vegetation data, obtained from IKONOS satellite images, and Digital Elevation Model obtained from digitized topographic maps. The study area is a northern slope of the Stanisoarei Mountains with a gradient of species from beech mixed and coniferous stands

**Keywords:** image segmentation, classification, GIS, DEM

## 1. Introduction

Remote sensing applications in the forest area are numerous as the information extracted from the satellite images can be useful in describing the forest stand and forest sites. In this general framework the large ecological studies are included, which observe the spatial distribution of some details of interest of the forest species on different geomorphological formations. Such a mapping of the species distribution can, of course, be done on the field by partial inventories and interpolations within some relatively homogenous populations. The advantage offered by an approach using of remote sensing is offered by the possibility of full inventory of the areas with certain characteristics. Although such a characterization of the forest stands is more general then the one inventoried on the field, the mapping accuracy can be seriously improved because of the current high quality recordings (Franklin 2001, Gao, 2009).

In this context is also included the distribution of some forest species on the geomorphological units and especially of the slopes. The aspect is attractive as the vegetation conditions are variable and impose a certain distribution of species, especially in the stands resulted from natural regeneration. The analysis possibilities are high, and the results are valuable, if we take into account the integration of the data resulted from satellite images classification with Geographical Informational Systems (GIS). Based on the satellite data there can be done rapidly and easily distribution maps which can comprise the information regarding certain geomorphological characteristics, even the meteorological and phenology data.

The objective of the study was to develop an interdisciplinary method to delineate tree species distribution in topographically-complex habitats.

## 2. Materials and methods

The study is located in the Vanatori Neamt National Park. As materials, we used pan-sharpened IKONOS 2 multispectral images and topographical plans (scale 1:5000) with contour lines.

For the satellite image classification we used field data regarding vegetation characteristics, composition, biometrical parameters. In order to choose the best classification there were compared two distinct work methods: one based on pixel classification (Pixel-Based Classification), done with ERDAS Imagine 9.1 and the second one based on multiresolution segmentation within the textural analysis program eCognition Professional 4.0. The classification precision was calculated in each case using field data, portions known form IKONOS satellite image and high resolution aerial photos. The vector data resulted from the classification allow the information integration in thorough ecological studies, in which geological and geomorphological aspects are to be taken into consideration. Keeping in mind that for the above situation presented above the geological characteristics are almost constant, there were studied the morphometric parameters, attempting their correlation to the vegetation characteristics, extracted from the available satellite images.

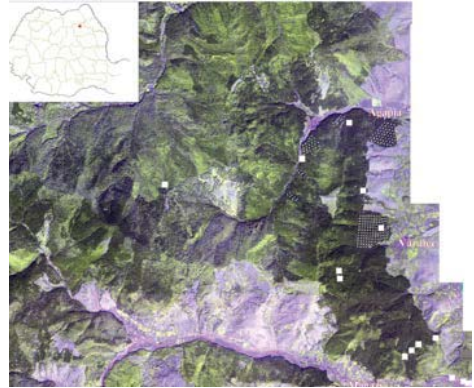

Fig. 1 Location of the study area

For the GIS information and remote sensing data integration it was imperative the designing of filed three-dimensional model based on the vectorised contour lines. As a cartographic source we used 1:5000 topographic plans with contour lines, which were turned into digital form with ArcGIS 9.3. For the interpolation Krigging method was used, and the terrain model was analysed in raster format (grid); the processing was done with the 3D Analyst module from ArGIS 9.3.

A brief interpretation of the geomorphological information and data related to the forest vegetation is represented by the designing of a three-dimensional image; it was done by correlating the image pixels flat position with the values corresponding to the three-dimensional terrain model, designed after the contour lines on the topographic plans using ArcScene from ArcGIS 9.3 (fig. 5). The utility of the application is more one of 3D visualising of general data, extended the observations at a large scale.

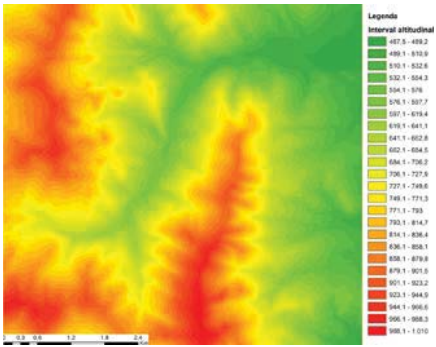

Fig. 2. 3D terrain model for an area within Production Unit III Agapia

Specific geomorphological parameters were extracted from the 3D terrain model using the GIS application specific modules which allowed the precise quantification and on large areas as it follows:

- Elevation can be extracted directly from the 3D terrain model, by identifying the areas of interest and applying them as a mask on the DEM.
- Slope is calculated for each triangle constituted in the elevation values interpolation process, designing work „facets" with constant slope.
- Aspect is associated to each facet as well as an orientation of the steepest slope line (ArcGIS Userguide).

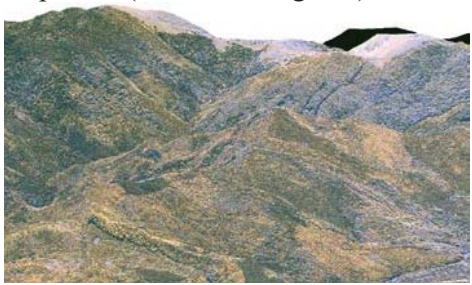

Fig. 3 IKONOS image overlay on terrain elevation model (vertical exaggeration 1.5 x)

In the case of each terrain model there were used interpolation algorithms. These, in vector format led to the projection of some triangles which cannot be overlapped with polygons designed by raster formats interpretation, as the satellite images are. In order to ensure compatibility of the data the specific files were transformed into raster files with the same resolution as IKONOS images used in Object Oriented Classification respectively 1m (Haralick, 1973, kayitakire, 2002).

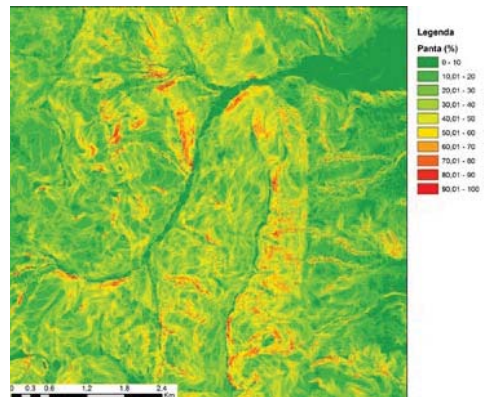

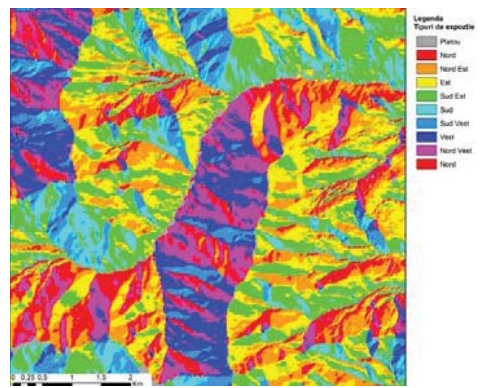

Fig.4. Slope and aspect of an area within the Production Unit III, Văratec Forest District

Even if the overlapping cannot be done by the real limits of the pixels, the methods based on the nearest neighbour identification can ensure a good overlapping precision. In order to exemplify the information usage mentioned in ecological studies a slope of Stanisoara Mountains was taken into consideration in the area Văratec – Agapia, slope with a shaded general exposition, extended on an altitudinal interval between 540 and 900 meters, with an average slope of approximately 25% (fig. 4).

The data regarding forest vegetation were taken from the vector file derived from the classified image segmentation. For the information representativity there were separated only the vectorial objects outlined as polygons corresponding to the portion of ridge taken into account. In this manner, the spruce plantations from the vicinity of the stream and the vegetation from the other slope were eliminated from the analysis.

Data overlay was based on the use of the same datum in the processing of georeferenced images (Stereo 1970, as defined in ArcGIS 9.3). IKONOS images georeferenced in the GCS WGS - 84 –UTM projection (Universal Transversal Mercator) were verified and then projected in the national datum - Krasovski ellipsoid and Stereo 70 projection. The digital terrain model was obtained from topographical plans with contour lines on a scale of 1:5000, by vectorisation and transformation using Topo to Raster function. For the overlay of vector and data extraction the Zonal Statistic (as a table) module of ArcGIS 9.3 was used.

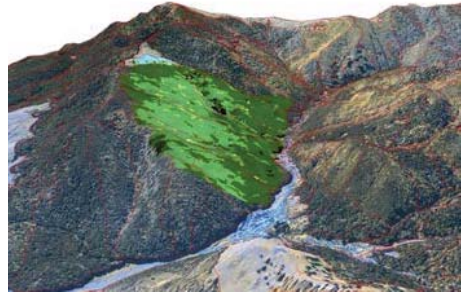

Fig. 5. Vegetation related classified image – study area for ecological application, in the topographical context of the Văratec area (vertical exaggeration 2x)

## 3. Results

By applying Zonal Statistic as a table function of ArcGIS a complex database was obtained, recording the computed geomorphological parameters for each exported polygon, originated from the segmentation and per object classification. The database can be interrogated in order to delineate the distribution of areas occupied by different vegetation classes in relation to the topographic parameters.

The distribution charts for the distribution of the identified groups of species on the slopes show that there is a distinct pattern of distribution of the species with specific requirements for ecological parameters (fig. 6):

Coniferous forests have higher proportions in the lower and upper part of the slope, between 540 and 700 m and between 800 and 900 m;

Areas classified as mixed forests with beech and coniferous are mainly present lower-middle part of the slope between 600 and 850 m, with a maximum around the elevation class 750 m;

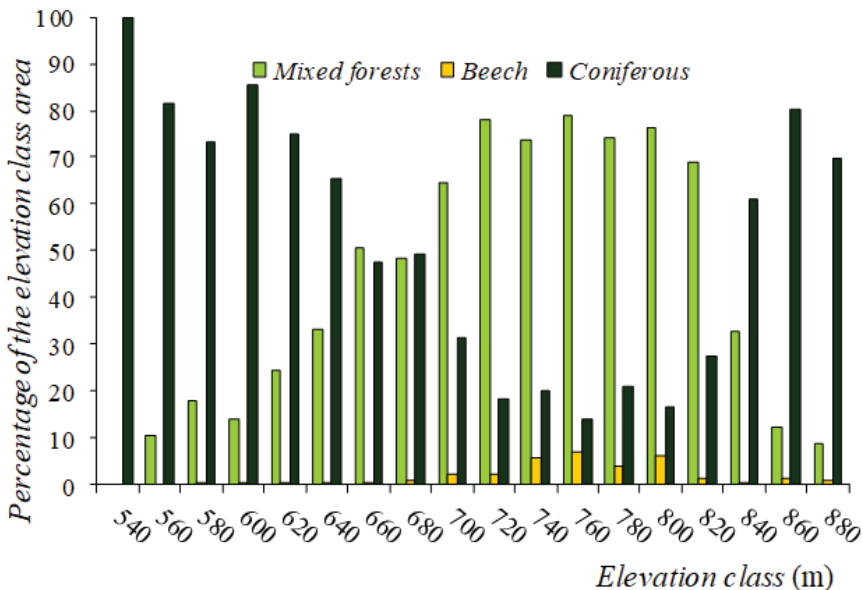

Fig. 5. Area percentage distribution in relation to the specie and the altitude class

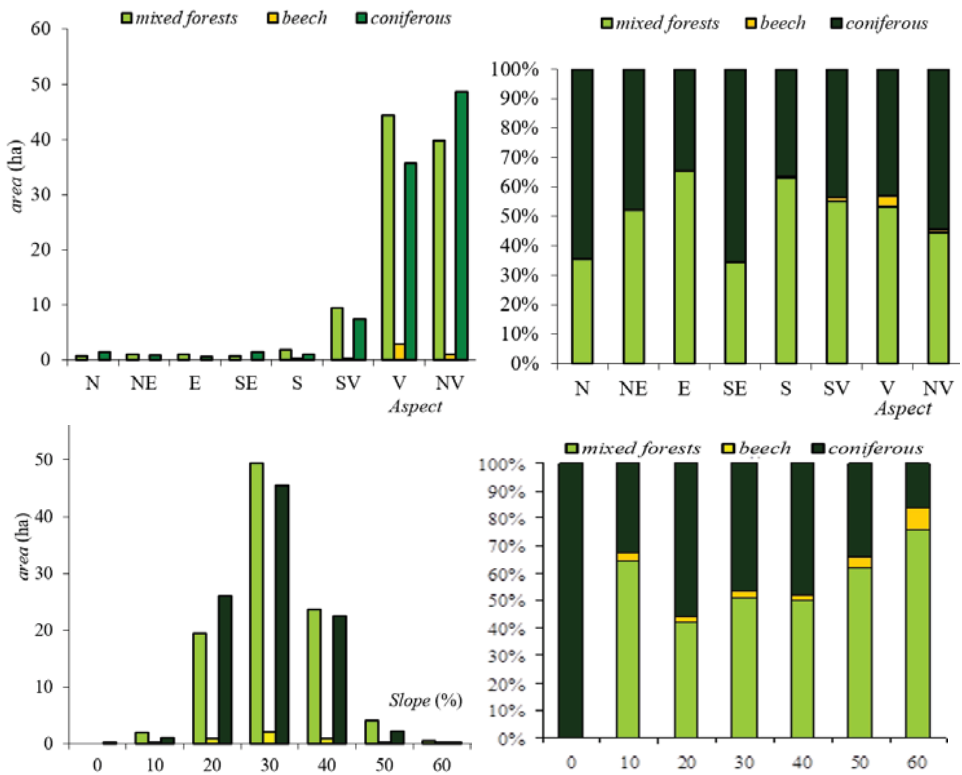

Fig. 6. Vegetation distribution on the slopes: a, b according to aspect; c,d according to declivity classes

Small portions of beech stands are located in the central part of the slope, on elevations between 700 and 800 m.

The alternant distribution of vegetation on the slope can be related to the temperature inversion that is a phenomenon characteristic for narrow valleys and predominant shaded slopes. The effect of the phenomenon is revealed by the natural regeneration of the stands on the entire slope, which can offer information on the specific requirements of the species involved.

In the lower part of the slope, exceptions from the natural distribution of species can be noticed: sycamore and scots pine plantations on areas less than 0.5 ha. During the segmentation and per object classification process, these areas have been included in the corresponding classes mixed forests and coniferous forests (which they actually represent).

The elevation related variation of stand composition can be correlated with respective climatic parameter modification, which can also be modelled through GIS analysis, based on recurring meteorological data. The chart information can be compared with the ecological distribution charts presented in the literature (Stănescu et al., 1997), which do not account for the inversion that can be present in such areas. Any comparison with such charts must also include a correction of the altitude with the values of the latitude, as the charts are constructed for the latitude of 45o (Florescu, Nicolescu, 1996).

The spatial distribution of species groups on different topography was obtained by GIS overlay of classified image objects on the surface model (fig. 3). The morphological parameters taken into account are the aspect and slope of each pixel, defined by comparison with the other pixels around (moving window). The analysis of the absolute values and percentages of each class showed that:

 Predominant slope aspect is North and Northwest; other aspect types are found on smaller areas and are related to the fragmentation of the slope due to the secondary valleys;

 Pure beech stands are found only in shaded parts of the slope; the other areas are covered with mixed forests and coniferous;

 The entire slope has a high declivity, which increase the effect of aspect on the forest vegetation.

The repartition on elevation, slope and aspect classes showed particular cases of species distribution interference, given by the different behaviours of species in regard to the values of the ecologic factors. Due to the temperature inversions given by the narrow valley bottom, the pure beech stands are found only in the middle of the slope, where the species found favourable environmental conditions (Stănescu et. al., 1997) and "wins" the interspecies competition. On higher elevations, the beech decreases in proportion due to the vertical temperature gradients.

The species composition modification is intensified by the increases in slope, which enhances the effect of the slope. The beech occupies sites with low variations in terms of aspect and slope, especially since the species is at the limit of the natural area.

*All papers in this publication were submitted on-line and have been edited to achieve a uniform format. Authors are responsible for the content of their paper.*